\documentclass[fleqn,12pt,twoside]{article}
\usepackage[headings]{espcrc1}

\readRCS
$Id: espcrc1.tex,v 1.2 2004/02/24 11:22:11 spepping Exp $
\usepackage{graphicx}
\usepackage[figuresright]{rotating}
\usepackage{amsmath,amssymb}

\newcommand{\tr}{\textnormal{Tr\,}}
\newcommand{\bea}{\begin{eqnarray}}
\newcommand{\eea}{\end{eqnarray}}
\newcommand{\be}{\begin{equation}}
\newcommand{\ee}{\end{equation}}

\title{
The boundary of the first order chiral phase transition in the
$m_\pi-m_K$--plane with a linear sigma model}

\author{T. Herpay
\address[egyik]{Department of Physics of Complex Systems,
E{\"o}tv{\"o}s University
}
\address[masik]{Research Group for Statistical Physics of the Hungarian
Academy of Sciences,\\ H-1117 Budapest, Hungary}
and
Zs. Sz{\'e}p 
\address{Research Institute for Solid State Physics and
	Optics of the Hungarian Academy of Sciences, H-1525 Budapest, Hungary}
	\thanks{Supported by OTKA Postdoctoral Grant no. PD 050015.}}
       
\runtitle{Phase boundary in the $(m_\pi~-~m_K)$--plane}
\runauthor{T. Herpay, Zs. Sz{\'e}p}

\begin{document}

\maketitle

\begin{abstract}
Tree-level and complete one-loop parametrisation of the linear sigma
model (L$\sigma$M) is performed and the phase boundary between first
order and crossover transition regions of the $m_\pi-m_K$--plane is
determined using the optimised perturbation theory (OPT) as a
resummation tool of perturbative series. Away from the physical point
the parameters of the model were determined by making use of chiral
perturbation theory (ChPT). The location of the phase boundary for
$m_\pi=m_K$ and of the tricritical point (TCP) on the $m_\pi=0$ were
estimated.
\end{abstract}

\section{Introduction}

Lattice studies show that at the physical point of the $m_{u,d}-m_s$--plane 
the transition as a function of the temperature is of crossover-type 
\cite{katz04}. In the $m_{u,d}=m_s=0$ limit effective models predict 
a first order phase transition and for $m_s=\infty$ and $m_{u,d}=0$ 
a second order phase transition \cite{pisarski84}. Theoretically it 
is interesting to know where the boundary of the first order region 
lies and where the TCP is located on the $m_u=0$ axis.

Several lattice studies with degenerate quarks
($m_{u,d}=m_s$) show that the critical pion mass on the boundary
decreases substantially, from $m_\pi^c\approx290$ MeV to 
$m_\pi^c=67(18)$ MeV, as finer lattices and improved actions are used 
(see {\it e.g.} \cite{Karsch_high_mpc,Karsch_low_mpc} ). A
low pion mass allows for studies in effective models based on the chiral
symmetry because it is expected that they work better 
closer to the chiral limit. Existing works performed in these
models give also a low value for the critical pion mass 
\cite{eff1,eff2,eff3}.  In
these works only the Goldstone masses were tuned, the other parameters of 
the Lagrangian were kept at their values determined at the physical point. 
We think that in this low mass regime effective model studies might
complement lattice investigations if coupling parameters can be
determined accurately. Hence we let the coupling constants to vary with
the pion and kaon masses when 
moving away from the physical point and determined their values using 
the low energy results of the ChPT, namely the
dependence of $f_\pi, f_K, m_\eta$ and $m_{\eta'}$ on $m_\pi$ and 
$m_K$. 

\section{The parametrisation of the linear sigma model}

The Lagrangian is written in terms of 
a $3\times 3$ 
matrix $M=\lambda_a(\sigma_a+i\pi_a)/\sqrt{2}$ 
($\lambda_a:a=1\dots 8$ are the Gell-Mann matrices and
$\lambda_0:=\sqrt{2/3} {\bf 1}$)
containing the nonet of scalar 
($\sigma_a$) and pseudoscalar ($\pi_a$) particles:
\bea
\nonumber
L(M)&=&\frac{1}{2}\tr(\partial_\mu M^{\dag} \partial^\mu M+\mu^2
M^{\dag} M)-f_1
\left( \tr(M^{\dag} M)\right)^2-f_2  \tr(M^{\dag}
M)^2\\
&&
-g\left(\det(M)+\det(M^{\dag})\right)+\epsilon_0\sigma_0+
\epsilon_8  \sigma_8.
\label{Lagrangian}
\eea
The determinant term breaks the axial $U(1)$ symmetry and
the external fields are introduced in order to give masses to the
pseudo-Goldstone bosons. From the original 0-8 basis one can switch with an
orthogonal transformation to a basis in which in the broken symmetry phase
the two non-vanishing expectation values are the non-strange 
($x=(\sqrt{2}\langle\sigma_0\rangle+\langle\sigma_8\rangle)/\sqrt{3}$) and 
the strange ($y=(\langle\sigma_0\rangle-\sqrt{2}
\langle\sigma_8\rangle)/\sqrt{3}$) condensates. 

\subsection{Tree-level parametrisation}

The tree-level parametrisation of the model can be performed at zero
temperature by solving a set of coupled linear equations
\cite{lenaghan,herpay1}. Only the trace of the mass matrix is used
in the mixing $\eta-\eta^\prime$ sector. Unfortunately the values the quartic 
coupling $f_1$ and the mass parameter $\mu^2$ can be obtained
using only the scalar sector which compared to the pseudoscalar sector 
is experimentally less known.  Besides, one knows nothing about the
$m_\pi$ and $m_K$-dependence of the scalar masses and as a consequence some
assumptions are needed. For example in \cite{herpay1} we required the
Gell-Mann-Okubo mass formula to be satisfied in the scalar sector on the
entire $m_\pi-m_K$--plane with the same accuracy as at the physical
point.
Performing a consistency check by comparing $m_\eta$ and $m_{\eta'}$ 
calculated in the L$\sigma$M
with their values given by the ChPT remarkable agreement was obtained 
up to a limiting value of $m_K\approx 800$ MeV, 
even away from the physical point.

\subsection{One-loop parametrisation}

The parametrisation at the one-loop level of the perturbation theory is more
complicated and there are many possible ways of selecting the 
set of non-linear equations. One such set of equations is presented
below (see \cite{herpay2} for details).

Because in the broken symmetry phase the tree-level mass can be
negative a resummation is needed. We performed this using the optimised
perturbation theory method of Ref. \cite{chiku98} where 
a  mass term is added to and subtracted from the Lagrangian, 
the difference between the original and the new mass is 
treated perturbatively (first at one-loop level). The new mass is
determined from the principle of minimal sensitivity which requires that the
one-loop pole mass for the pion ($M_\pi$) stays equal to its tree-level 
value ($m_\pi$).
This can be translated into a self-consistent equation for the pion
mass
\be
m_\pi^2=-\mu^2+(4f_1+2f_2)x^2+4f_1y^2+2gy+\textrm{Re}
\Sigma_\pi(p^2=m_\pi^2).
\ee
Two more equations are the one-loop pole mass of the kaon and
$\eta$ (smaller mass eigenvalue in the mixing sector): 
\bea
M_K^2=-\mu^2+2(2f_1+f_2)(x^2+y^2)+2f_2y^2-\sqrt{2}x(2f_2y-g)+
\textrm{Re} \Sigma_K(p^2=M_K^2),\\
\textnormal{Det}\left.
\begin{pmatrix}
p^2-m_{\eta_{xx}}^2-\Sigma_{\eta_{xx}}(p^2) \,\,&
-m_{\eta_{xy}}^2-\Sigma_{\eta_{xy}}(p^2) \\
-m_{\eta_{xy}}^2-\Sigma_{\eta_{xy}}(p^2) &
p^2-m_{\eta_{yy}}^2-\Sigma_{\eta_{yy}}(p^2) \,\,
\end{pmatrix}
\right|_{p^2=M_\eta^2}=0.
\eea
We require additionally the equality of the one-loop an tree-level kaon masses:
\be
M_K^2\overset{\displaystyle}{=}m_K^2=m_\pi^2-2gy+4f_2y^2-\sqrt{2}x(2f_2y-g),
\ee
and also make use of the PCAC relations:
\be\qquad
f_\pi=Z_\pi^{-\frac{1}{2}} M_\pi^{-2}(-iD^{-1}_\pi(p=0))x,
\label{PCAC_pi}
\qquad
f_K=Z_K^{-\frac{1}{2}} M_K^{-2}(-iD^{-1}_K(p=0))(x+\sqrt{2}y)/2.
\ee

Away from the physical point we
use for the determination of $m_\eta, f_\pi$ and $f_K$ the
formulae of the SU(3) ChPT in the large-$N_c$ limit.
Because in the one-loop parametrisation the renormalisation scale ($l$)
appears as a new parameter we checked how the other parameters and 
masses depends on it. The wave function renormalisation constants
have a plateau and the variation of the one-loop masses is the mildest in the
range $l\in (1000,1400)$ MeV, which was selected to be used for the 
determination of the phase boundary.

\section{Thermodynamics of the L$\sigma$M}

The order of the phase transition is determined by solving 
simultaneously the self-consistent gap equation for the pion mass
\be
m_\pi^2=-\mu^2+(4f_1+2f_2)x^2+4f_1y^2+2gy+
\Sigma_\pi(p^2=m_\pi^2,m_i(m_\pi),l),
\ee
where $\Sigma_\pi(p,m_i(m_\pi),l)$ is the pion self-energy,
and the two equations of state
\bea
\epsilon_x&=&-\mu^2x+2gxy+4f_1xy^2+
(4f_1+2f_2)x^3+
\textstyle \sum_i J_i t^x_i(x,y)
\textnormal{I}_{\textrm{tp}}[m_i(m_\pi),T],
\nonumber
\\
\epsilon_y&=&-\mu^2y+2gx^2+4f_1x^2y+(4f_1+4f_2)y^3+
\textstyle \sum_i J_i t^y_i (x,y)
\textnormal{I}_{\textrm{tp}}[m_i(m_\pi),T]\, .
\eea
The isospin multiplicity factor $J_i$ and the coefficients $t_i^x$ and 
$t_i^y$ are given in Ref. \cite{herpay1}.

The difference between the tree-level and one-loop parametrisation is that
in the first case one are enforced to throw out entirely the vacuum
fluctuations and use only the temperature dependent part of the
integrals, while in the second case the renormalised vacuum fluctuations are
taken into account. The difference between the two methods
of solving the model was checked quantitatively in \cite{herpay2}. 


\section{The phase-boundary in the $m_\pi-m_K$--plane}

In the case of the tree-level parametrisation the result on the phase
boundary is not too conclusive because of the dependence of the phase
boundary on the assumptions made for the scalar sector in the
process of parametrisation. We have not seen the trace of the tricritical
point. We estimated that the phase boundary crosses the diagonal of the
$m_\pi-m_K$-plane for critical pion mass in the range $m_\pi^c=40\pm 20$
MeV.

\begin{figure}[!t]
\centering{
\includegraphics[width=25pc, keepaspectratio]{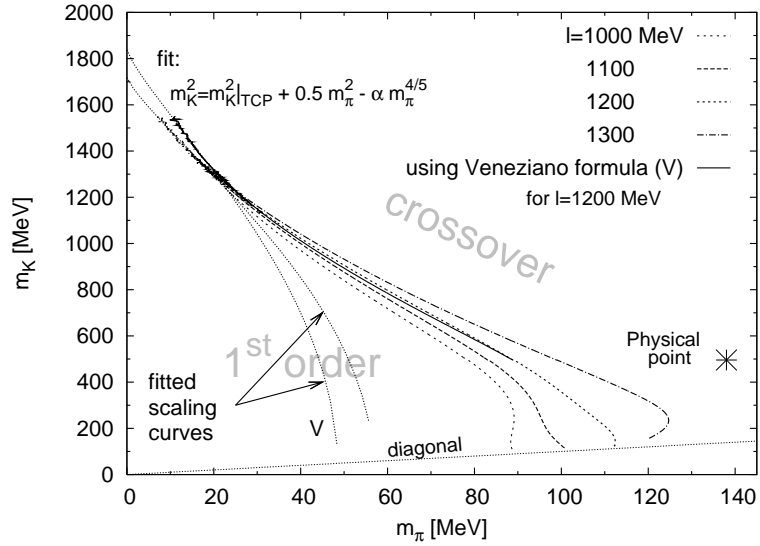}}
\caption{Phase boundary of the L$\sigma$M obtained using a one-loop
parametrisation at $T=0$ and the formulae of large $N_c$ ChPT for
continuation on the $m_\pi-m_K$--plane. 
}
\label{fig:boundary}
\end{figure}

As for the one-loop parametrisation, along the diagonal of the pion-kaon mass
plane we estimate $m_\pi^c=110\pm20$ MeV (see Fig.~\ref{fig:boundary}). 
At high values of $m_K$ the phase boundary is not 
renormalisation scale dependent. We can clearly see the scaling region of the trictitical point as
well. Unfortunately we were not able to locate it directly because the model
parametrisation breaks down just before arriving at the TCP along the
$m_\pi=0$ line, but based on the scaling near TCP we estimate it to be in
the interval $m_K^{TCP}\in\{1700,1850\}$ MeV.
We can directly compare our result on the location of the TCP with the recent 
lattice result of Ref. \cite{philipsen} by showing the phase boundary 
in the quark mass-plane. In Ref. \cite{philipsen} the physical point is more 
close to the boundary than in our case and the value of the mass of the 
strange quark at the TCP is $m_s^\textnormal{TCP}=3 m_s$ 
($m_s^\textnormal{phys}$ is the value of the strange quark mass at the 
physical point). In our case 
$m_s^\textnormal{TCP}=13-15 \times m_s^\textnormal{phys}$. 
In Ref. \cite{philipsen} the location of the TCP was estimated also 
by fitting with the corresponding scaling equation 
($m_{u,d}\approx (m_s^\textnormal{TCP}-m_s)^{5/2}$), but only
points with $m_s\le m_s^\textnormal{phys}$. The investigation performed 
in the L$\sigma$M shows that the scaling region sets in very close 
to the $m_u=0$ ($m_\pi=0$) axis. It would be interesting to know how 
the lattice estimate would change if points more closer to the $m_u=0$ 
axis were available.

\end{document}